\documentclass[twocolumn]{svjour3}

\makeatletter
\DeclareRobustCommand*\cal{\@fontswitch\relax\mathcal}
\makeatother

\usepackage[utf8]{inputenc}
\usepackage[T1]{fontenc}

\usepackage{amssymb}
\usepackage[switch]{lineno}

\usepackage{amsmath}
\usepackage{stmaryrd}
\usepackage[ruled,vlined]{algorithm2e}
\usepackage{url}
\usepackage{paralist}
\usepackage[hidelinks]{hyperref}
\hypersetup{bookmarksdepth=3,bookmarksopen,bookmarksnumbered} \usepackage[numbers,sort&compress]{natbib}
\usepackage{todonotes}
\usepackage{datetime2}
\usepackage{xstring}
\usepackage{subcaption}
\usepackage{booktabs}
\usepackage{colortbl}
\usepackage{catchfile}

\makeatletter
\def\cl@chapter{\@elt {theore}}
\makeatother
\usepackage[capitalise]{cleveref}

\usepackage{placeins}
\usepackage{mdframed}
\usepackage{mathabx}

\definecolor{codehighlight}{RGB}{220, 255, 220}
\definecolor{operator}{RGB}{0, 102, 187}

\usepackage{xspace}
\usepackage{bm}
\usepackage{pifont}

\usepackage{pgfplots}
\usepgfplotslibrary{groupplots}
\usepackage{pgfplotstable}

\newcommand{\Z}{\mathbb{Z}}

\newcommand{\sem}[1]{\llbracket#1\rrbracket}
\newcommand{\ignore}[1]{}
\newcommand{\angl}[1]{\langle#1\rangle}

\newcommand\restr[2]{{\left.\kern-\nulldelimiterspace#1\vphantom{|}\right|_{#2}}}
\newcommand\restrs[2]{{\left.\kern-\nulldelimiterspace#1\vphantom{|}\right|^\sharp_{#2}}}
\makeatletter
\newcommand*{\declarecommand}{  \@star@or@long\declare@command
}
\newcommand*{\declare@command}[1]{  \provide@command{#1}{}    \renew@command{#1}}
\makeatother

\newcommand{\N}{{\cal N}}

\newcommand{\B}{{\cal B}}
\newcommand{\X}{{\cal X}}

\newcommand{\R}{{\cal R}}
\newcommand{\U}{{\cal U}}

\declarecommand{\G}{{\cal G}}
\newcommand{\D}{{\cal D}}
\newcommand{\Vars}{{\cal X}}
	\declarecommand{\C}{{\cal C}}

\renewcommand{\C}{{\cal C}}

\newcommand{\twoClusters}{[\X]_2}
\makeatletter\@ifclassloaded{llncs}  {}  {}\makeatother\makeatletter\@ifclassloaded{llncs}  {}  {}\makeatother

\makeatletter\@ifclassloaded{llncs}  {}  {}\makeatother\newcommand{\aSemR}[1]{\llbracket#1\rrbracket^\sharp}

\begin{document}

\title{Non-Numerical Weakly Relational Domains}
\author{
     Helmut Seidl\and
     Julian Erhard\and
     Sarah Tilscher\and
     Michael Schwarz
}
\authorrunning{H. Seidl et al.\ }

\institute{
	Helmut Seidl \and Julian Erhard \and Sarah Tilscher \and Michael Schwarz
	\at Technische Universit\"at M\"unchen, Garching, Germany\\
	\email{\{helmut.seidl, julian.erhard, sarah.tilscher, m.schwarz\}@tum.de}
}
\date{\phantom{\today}}
\journalname{}

\maketitle
% \linenumbers
\begin{abstract}
	The weakly relational domain of \emph{Octagons} offers a decent compromise between precision and efficiency
	for numerical properties.
		Here, we are concerned with the construction of non-numerical relational domains.
	We provide a general construction of weakly relational domains, which we
	exemplify with an extension of constant propagation by disjunctions.
		Since for the resulting domain of 2-disjunctive formulas, satisfiability is NP-complete, we provide
	a general construction for a further, more abstract weakly relational domain where the abstract operations of restriction
		and least upper bound
	can be efficiently implemented.

	In the second step, we consider a relational domain that tracks conjunctions of inequalities between variables,
	and between variables and constants for arbitrary partial orders of values.
		Examples are sub(multi)sets, as well as prefix, substring or scattered substring orderings on strings.
		When the partial order is
		a lattice, we provide precise polynomial algorithms for satisfiability, restriction, and
	the best abstraction of disjunction. 		Complementary to the constructions for lattices, we find that, in general, satisfiability of conjunctions is NP-complete.
	We therefore again provide polynomial abstract versions of restriction, conjunction, and join.
		By using our generic constructions,
	these domains are extended to weakly relational domains that additionally track \emph{disjunctions}.

	For all our domains, we indicate how abstract transformers for assignments and guards can be constructed.

\end{abstract}

\section{Introduction}\label{s:intro}
	Relational analyses have been observed to be indispensable for verifying intricate
	program properties. In particular, this is the case when for the purpose of verification,
	\emph{ghost} variables have been introduced which must be related to program variables.
		Termination may be verified by introducing a \emph{ghost} loop counter, which can be proven bounded by a relational domain
	relating it to the actual bounded iteration variable \cite{Albert2014}.
		The validity of string operations on null-terminated strings as employed, e.g., in the programming language C,
	may be verified by introducing \emph{ghost} variables for the length of a buffer
	as well as for tracking the position of the null byte in the buffer~\cite{Sagiv2001}.
		It also has been observed that
	\emph{monolithic} relational domains such as the polyhedra abstract domain \cite{Cousot78}
	scale badly to larger programs.
			Therefore, \emph{weakly relational} domains have been proposed which can only express simple relational properties,
	but have the potential to scale better \cite{Mine04}.
		Examples of weakly relational \emph{numerical} properties are
    	the  \emph{Two Variables Per Inequality} domain \cite{Simon02},
	or domains given by a finite set of \emph{linear templates} \cite{Sankaranarayanan05}.
	    		The most prominent example of a template numerical domain is the \emph{Octagon} domain \cite{Mine01,Mine06Oct}
	which allows tracking upper and lower bounds not only of program variables but also of sums and differences of
	\emph{two} program variables.
		One such octagon abstract relation  could, e.g., be given by the conjunction
	\[
		(-x\leq -5)\wedge (x\leq 10)\wedge(x+y\leq 0)\wedge(x-z\leq 1)
	\]
        	\emph{Octagons} thus can be considered as a mild extension of the non-relational domain of
	\emph{Intervals} for program variables, and a variety of efficient algorithms have been provided
	\cite{Bagnara08,Bagnara09,King,Schwarz23b}.
	Here, we are concerned with constructing \emph{non-numerical} abstract domains.

	For that, we provide a general technique to construct from \emph{every} relational domain
	a weakly relational domain.
		As one instance of the general construction, we consider
	2-dis\-junc\-tive constants as mentioned in \cite{Schwarz23}.
	This weakly relational domain allows, e.g., to relate the names of functions with function pointers as in
	the formula
	\[
	 x=\textsf{"foo"}\wedge y=\&\textsf{foo}\;\vee\;
	 x=\textsf{"bar"}\wedge y=\&\textsf{bar}
	 \]
				Since satisfiability of formulas from that domain turns out to be NP-complete,
	we provide a further mild abstraction, again for arbitrary relational domains,
	to provide us with a weakly relational domain where all required operations become tractable.

	Another family of relational non-numerical domains has been introduced by \citet{Arceri22}.
	Based on a partial order
	of values, conjunctions of ordering constraints $x\sqsubseteq y$ for program variables $x,y$ are considered.
		They observe that analyses of prefixes or the substring relation could be helpful
	for programs in programming languages supporting high-level operations on strings.
		Here, we study this kind of directed domains in greater detail.
		For conjunctions of inequalities over some partial order $P$, we extend the constraints from \citet{Arceri22}
	by allowing for variables both lower and upper bounds from $P$.
		For arbitrary partial orders, though, we find that then satisfiability is NP-complete.
	Partial orders $p$ that are lattices form a notable exception.
		An instance of this are subsets of some universe or multisets.
			For lattices, we show that satisfiability is decidable in polynomial time.
	Moreover, we provide polynomial constructions both for restriction as well as the
	optimal join operation.
		Turning to general partial orders of values, we thus cannot hope for polynomial algorithms.
	Therefore, we provide a meaningful abstraction so that both abstract restriction as well as join
	is again polynomial.
		This family of relational domains is already weakly relational.
	Still, our generic constructions can be applied to obtain
	more expressive weakly relational domains that additionally support disjunctions at a limited amount of
	extra costs.

The paper is organized as follows:
\cref{s:domain} provides background definitions on relational domains. It formally introduces our notion of
weakly relational domains and provides a general construction of weakly relational domains.
\cref{s:const} is dedicated to disjunctive constants.
When applying the generic construction from the last section to this relational domain,
the weakly relational domain of 2-disjunctive constants is obtained.
Here, we prove that satisfiability for these formulas still is NP-complete.
Therefore, a generic abstraction technique
is presented so that,
when applied to disjunctive constants,
normalization, projection, as well as least upper bounds all turn out to be polynomial time.

Finally, abstract transformers for assignments as well as guards are derived.
\cref{s:directed} then introduces \emph{directed domains} which do not track equalities but inequalities over
a partial order of values.
While the first subsection provides polynomial constructions for the case that the partial order for values
is a lattice,
the second subsection is concerned with arbitrary partial orders as value domain.
Since satisfiability, in general, turns out to be NP-complete, again a polynomial abstraction is provided.
In a further subsection, we indicate how the generic constructions from the last sections provide us with weakly relational
domains that additionally support disjunctions of inequalities. We exemplify the resulting domains
with conjunctions and disjunctions of inequalities over the integers.
In the final subsection, dedicated abstract transformers are constructed for assignments, while the last subsection
discusses the treatment of guards.
\cref{s:conclusion} summarizes the contributions and sketches further directions of research.

\section{Weakly Relational Domains}\label{s:relational}\label{s:domain}

Let us recall basic definitions for relational domains.
We mostly follow the notation used in previous work~\cite{Schwarz23}, where the notion of
$2$-decomposability has been introduced.
Let $\Vars$ be some finite set of variables.
A \emph{relational domain} $\R$
maintains relations between variables in $\Vars$. We require that a relational domain is a bounded lattice, i.e., has a partial order $\sqsubseteq$, a least element $\bot$, a greatest element $\top$, as well as binary operators for the greatest lower bound  (meet)~$\sqcap$ and the least upper bound (join)~$\sqcup$.
We do not demand relational domains to be \emph{complete} lattices, i.e., to provide for \emph{every} subset of elements a
least upper bound: the polyhedral domain, e.g., is not complete \cite{Cousot78}.
However, we demand that a relational domain supports the following monotonic operations:
\[
\begin{array}{rcl}
\aSemR{x\,{:=}\,e}&:&\R\to\R\text{ (assignment of $e$ to $x$)}	\\
\restr{\cdot}{Y} &:&\R\to\R\text{ (restriction to $Y \subseteq \Vars$)} \\
\aSemR{?c}&:&\R\to\R\text{ (guard for condition $c$)}
\end{array}
\]
where $e$ and $c$ are from some expression and condition language, respectively.

\medskip

\noindent
The abstract transformers for basic actions of programs are given by these functions.
Restricting a relation $r$ to a subset $Y$ of variables amounts to \emph{forgetting} all information about variables in
$\Vars \setminus Y$.
Thus, we require that
\begin{equation}
	\begin{array}{lll}
		\restr{r}{\Vars}	&=&	r \\
		\restr{r}{\emptyset}	&=&	\left\{
			\begin{array}{ll}
			\bot&\text{if}\;r=\bot	\\
			\top&\text{otherwise}
			\end{array}\right.	\\
		\restr{r}{Y_1} 	&\sqsupseteq& \restr{r}{Y_2}\qquad\text{when}\; Y_1 \subseteq Y_2	\\
		\restr{(\restr{r}{Y_1})}{Y_2} &=& \restr{r}{Y_1 \cap Y_2}
	\end{array}
	\label{d:restrict}
\end{equation}
A restriction $\restr{\cdot}{Y}$ to some set $Y$ therefore is an \emph{idempotent} operation.
We remark that from these axioms it follows that $\restr{\bot}{Y} = \bot$ and $\restr{\top}{Y} = \top$ for any $Y \subseteq \X$.
Given that there is some relation $r_c\in\R$ describing all states satisfying the condition $c$,
the transformation for the guard $?c$ can be described by
\begin{equation}
	\aSemR{?c} r = r\sqcap r_c
	\label{d:guard}
\end{equation}
-- at least, if there is a concretization function $\gamma$ such that
\begin{equation}
\gamma\,(r_1\sqcap r_2) = \gamma\,r_1 \cap \gamma\,r_2
\label{eq:gamma}
\end{equation}
i.e., the binary meet operation is \emph{precise}.
\begin{example}\label{e:rel}
For numerical variables, a variety of such relational domains have been proposed, e.g.,
(conjunctions of) \emph{affine equalities} \cite{Karr1976,SeidlMMO04,SeidlMMO07} or
\emph{affine inequalities} \cite{Cousot78}.
	For affine equalities or inequalities,
restriction to a subset of $Y$ of variables corresponds to the geometric projection onto the
subspace defined by $Y$, combined with arbitrary values for variables $z\not\in Y$.
\qed
\end{example}
\noindent
One way to tackle the high cost of relational domains is to track the relationships not between all variables,
but only between \emph{subclusters} of variables. We call such domains \emph{Weakly Relational Domains}.

For a subset $Y\subseteq\Vars$, let $\R^{Y} = \{\restr{r}{Y} \mid r\in\R \}$ be the set of all abstract values from $\R$
that contains only information on those variables in $Y$.
For any collection ${\cal S}\subseteq 2^{\Vars}$ of \emph{clusters} of variables, a relation $r\in\R$ can be
\emph{approximated} by a meet of
relations from $\R^Y, Y\in\cal S$ since for every $r\in\R$,
\begin{equation}
r\sqsubseteq \bigsqcap_{Y\in\mathcal{S}}\restr{r}{Y}
\label{def:cluster-approx}
\end{equation}
holds, as $r \sqsubseteq \restr{r}{Y}$ holds for each $Y \in S$.
In fact, the right-hand side of \eqref{def:cluster-approx} is the \emph{best} approximation of $r$ by some meet over
abstract relations $s_Y,Y\in{\cal S},$ with  $s_Y\in\R^Y$, i.e., with $\restr{s_Y}{Y} = s_Y$, since
\[
\begin{array}{lcl}
\restr{r}{Y}	&\sqsubseteq&	\restr{(\bigsqcap{Y'\in{\cal S}} s_{Y'})}{Y}	\\
		&\sqsubseteq&	\restr{s_Y}{Y}	\qquad\qquad\quad\text{(by monotonicity of restriction)}	\\
		&=& 		s_Y
		\end{array}
\]
holds for all $Y\in{\cal S}$.

\citet{Schwarz23} have introduced $2$-\emph{decomposable} relational domains. These are domains where
the full value $r$ can be recovered from the restrictions of $r$ to all clusters $p$ from the set
$\mathcal{S}=\twoClusters$ of non-empty clusters of variables of size at most $2$.
Furthermore, \citet{Schwarz23} ask for binary least upper bounds to be determined by computing
within these clusters only. More precisely, this amounts to requiring the following two properties
\begin{eqnarray}
    r =& \bigsqcap_{p\in\twoClusters} \restr{r}{p}
	\label{def:decomp1}	\\
  \restr{\left(r_1\sqcup r_2\right)}{p} =&	\restr{r_1}{p}\sqcup \restr{r_2}{p}\qquad(p\in\twoClusters)
	\label{def:decomp2}
    \end{eqnarray}
to hold for all abstract relations $r,r_1,r_2\in\R$.
The most prominent example of a $2$-decomposable domain is the
\emph{octagon} domain \cite{Mine01} -- either over rationals or integers, while
\emph{affine equalities}
or \emph{affine inequalities}
are examples of domains that are not $2$-decomposable.

\emph{Any} relational domain $\R$, however, which satisfies \eqref{def:decomp2}
gives rise to a 2-decomposable domain $\R_2$ of its 2-cluster approximations.

For $r\in\R$, let $\overline{r} =\bigsqcap_{p\in\twoClusters}\restr{r}{p}$ denote the approximation of $r$
by the meet of its restrictions to clusters $p\in\twoClusters$.
Let $\R_2$ denote the subset of $\R$ of all abstract relations of the form $\overline{r},r\in\R$,
where the ordering is inherited from $\R$. In particular, $\bot$ as well as $\top$ from $\R$ are also in $\R_2$.

\begin{theorem}\label{t:2-decomp}
Assume that $\R$ is an abstract relational domain which satisfies \eqref{def:decomp2}.
Then the following holds:
\begin{enumerate}
\item	$r = \overline{r}$ for all conjunctions $r=\bigsqcap_{p\in\twoClusters}s_p$ with
	$s_p\in\R^p, p\in\twoClusters$, i.e., all such conjunctions are contained in $\R_2$.
	\item	For $r_1,r_2\in\R_2$, the abstract relation $r_1\sqcap r_2$, as provided by $\R$, is in $\R_2$.
\item	The binary least upper bound operation in $\R_2$ exists and is given by
	\[
	r_1\sqcup_2 r_2 = \bigsqcap_{p\in\twoClusters} (\restr{r_1}{p}\sqcup\restr{r_2}{p})
	\]
\item	For $\R_2$, the best approximation $\restr{r}{Y,2}$ to the restriction $\restr{r}{Y}$ of $r\in\R_2$
	onto some subset $Y\subseteq\X$ of variables is given by
	\[
	\restr{r}{Y,2}	= \bigsqcap_{p\in\twoClusters} \restr{r}{p\cap Y}
	\]
\item	the partial order $\R_2$ with the given binary greatest lower and least upper bounds is a 2-decomposable
	relational domain.
	\end{enumerate}
\end{theorem}

\begin{proof}
For a proof of statement (1), we first observe that for each $p\in\twoClusters$,
\[
\restr{r}{p} =
\restr{\left(\bigsqcap_{p\in\twoClusters} s_p\right)}{p}
\sqsubseteq\restr{s_p}{p} = s_p
\]
by monotonicity and idempotence of restriction.
Thus,
\[
r \sqsubseteq \overline{r} =
\bigsqcap_{p\in\twoClusters} \restr{r}{p}\sqsubseteq
\bigsqcap_{p\in\twoClusters} s_p = r
\]
where the first inequality follows from \cref{def:cluster-approx}. Thus, statement (1) follows.

For a proof of statement (2), consider elements $r,s\in\R_2$. Then
\[
r\sqcap s =
\bigsqcap_{p\in\twoClusters} \restr{r}{p}\sqcap
\bigsqcap_{p\in\twoClusters} \restr{s}{p}
=
\bigsqcap_{p\in\twoClusters} (\restr{r}{p}\sqcap\restr{s}{p})
\]
Now, we claim that for every $p\in\twoClusters$,
\[
\restr{r}{p} \sqcap \restr{s}{p}	=\restr{(\restr{r}{p}\sqcap \restr{s}{p})}{p}
\]
To prove the claim, we argue that
\[
\begin{array}{lcl@{\quad}l}
\restr{r}{p} \sqcap \restr{s}{p}
	&\sqsubseteq&
	\restr{(\restr{r}{p}\sqcap \restr{s}{p})}{p}	& \text{(by monotonicity)}	\\
	&\sqsubseteq&
	\restr{(\restr{r}{p})}{p}\sqcap \restr{(\restr{s}{p})}{p} &\text{(by monotonicity)}	\\
	&=&	\restr{r}{p}\sqcap\restr{s}{p}	&\text{(by idempotence)}
\end{array}
\]
and the claim follows.
So far, we have proven that
\[
r\sqcap s = \bigsqcap_{p\in\twoClusters} t_p
\]
for some $t_p\in\R^p$, $p\in\twoClusters$. Then, statement (2) follows from statement (1).

For a proof of statement (3), we note that any upper bound of $r_1,r_2$ in $\R_2$ is also an upper bound of $r_1\sqcup r_2$
in $\R$. Therefore, the \emph{least} upper bound od $r_1,r_2$ in $\R_2$ is given by $\overline{r_1\sqcup r_2}$.
We calculate:
\[
\begin{array}{lll@{\quad}l}
	\overline{r_1\sqcup r_2}
	&=&
	\bigsqcap_{p\in\twoClusters} \restr{(r_1\sqcup r_2)}{p}
		&\text{(by definition)}\\
	&=&
	\bigsqcap_{p\in\twoClusters} (\restr{r_1}{p}\sqcup \restr{r_2}{p})
		&\text{(by \eqref{def:decomp2})}
\end{array}
\]
and statement (3) follows.

The best approximation of $\restr{r}{Y}$ in $\R_2$
is given by $\overline{\restr{r}{Y}}$. Thus, we have
\[
\restr{r}{Y,2}
= \bigsqcap_{p\in\twoClusters} \restr{(\restr{r}{Y})}{p}
= \bigsqcap_{p\in\twoClusters} \restr{r}{Y\cap p}
= \bigsqcap_{p\in\twoClusters} \restr{(\restr{r}{p})}{Y}
\]
i.e., it can be determined by applying the restriction onto variables from $Y$ for each cluster $p\in\twoClusters$
separately. This implies statement (4).

Statement (5) is an immediate consequence of
statements (3) and (4).
\qed
\end{proof}

\noindent
	The polyhedral domain, e.g., satisfies \eqref{def:decomp2}.
	Applied to the polyhedral relational domain, the construction from \cref{t:2-decomp} results in the
	domain of affine inequalities with at most two variables per inequality
	\cite{Simon02}.

According to \cref{t:2-decomp}, every value $r$ from the $2$-decomposable relational domain $\R_2$ can be represented as the meet of its restrictions to
$2$-clusters, i.e., by the collection $\langle\restr{r}{p}\rangle_{p\in  \twoClusters}$.
We call this representation \emph{normal}, and an algorithm that computes it \emph{normalization}.
Consider now an \emph{arbitrary} collection $\angl{s_p}_{p\in \twoClusters}$ with $s_p\in\R^p$ with
$r=\bigsqcap_{p\in\twoClusters}s_p$.
Then $\restr{r}{p} \sqsubseteq s_p$ always holds, while equality need not hold.
In the \emph{Octagon} domain over the rationals or the integers,
the normal representation of an octagon value corresponds to its
\emph{closure} as introduced in previous work \cite{Mine01,Bagnara08}.
While for rational Octagons, closure in cubic time was already proposed by \citet{Mine01},
it is much more recent that
a corresponding algorithm was provided for the case when
constraints are interpreted over integers \cite{Bagnara08,Bagnara09}.

Subsequently, we introduce non-numerical weakly relational domains and provide polynomial algorithms for these.

\section{Disjunctive Constants}\label{s:const}

Constant propagation relies on a domain that maintains conjunctions of atomic propositions
$x=a$ where $x$ is a program variable and $a$ is from a finite set $U$ of possible values.
In the following, we consider a (mild) generalization of this domain where also \emph{disjunctions} of at most two
atomic propositions are allowed.

Assume we are given a finite set $U$ representing possible values for variables from $\Vars$.
We consider propositions of the form $(x\in A)$ for $A\subseteq U$ which correspond
to the disjunction of atomic propositions $x=a,a\in A$.
Thus, the proposition $x\in A$ for some $A\subseteq U$ can be understood as an atomic proposition of a
\emph{multi-valued} propositional logic
where $A$ serves as the set of logical values of the propositional variable $x$
\cite{Beckert2000}.
Every monotonic Boolean combination $\Psi$ of propositions $x\in A$ with $x\in\X, A\subseteq U$,
represents a function $\sem{\Psi}:(\Vars\to U)\to\B$
defined by
\[
\begin{array}{lll}
\sem{x\in A}\;\sigma	&=& 	(\sigma\,x)\in A	\\
\sem{\Psi_1\vee\Psi_2}\;\sigma	&=&
	\sem{\Psi_1}\,\sigma\vee\sem{\Psi_2}\,\sigma	\\
\sem{\Psi_1\wedge\Psi_2}\;\sigma	&=&
	\sem{\Psi_1}\,\sigma\wedge\sem{\Psi_2}\,\sigma	\\
\end{array}
\]
Let $\C[U]$ denote the complete lattice of all equivalence classes of formulas $\Psi$
where the ordering is semantic implication. The least element in this ordering can be represented by the
empty disjunction or $\bot$ (\emph{false}), while the greatest element is equivalent to
the empty conjunction or $\top$ (\emph{true}).
Each formula $\Psi$ has an equivalent CNF as well as an equivalent DNF
where each clause (conjunction) contains at most one proposition $x\in A$
for every variable $x$.
Converting $\Psi$ into DNF allows checking satisfiability and
computing the restriction $\restr{\Psi}{Y}$ onto a subset $Y\subseteq\X$ of variables.
A formula for $\restr{\Psi}{Y}$ is obtained from a DNF for $\Psi$
where each conjunction contains
at most one proposition for each variable by the following steps:
First, every conjunction which contains $y\in\emptyset$ for some $y$ is removed.
From each remaining conjunction, then
every proposition $y\in A$ with $y\not\in Y$ is removed.
It follows that $\restr{\Psi}{Y}$ is distributive, i.e., commutes with binary least upper bounds.

For an arbitrary $\Psi\in\C[U]$, computing an equivalent DNF is an exponential time operation.
The same holds if all restrictions $\restr{\Psi}{\{x,y\}}$ are computed via this normal form.
Let $\C_2[U]$ denote the 2-decomposable domain obtained from $\C[U]$ according to theorem \ref{t:2-decomp}.
The lattice $\C_2[U]$ consists
of all elements $\Psi$ which can be represented
as conjunctions of clauses with at most two propositions $x\in A_x$ per clause.
According to theorem \ref{t:2-decomp}, the least upper bound operation $\sqcup_2$ for $\C_2[U]$
can be realized by a clusterwise disjunction.
In particular, it does \emph{not} coincide with logical disjunction
-- but is an over-approximation of it.

\begin{example}
	Let $\Psi_1 \equiv (x\in\{a\})$ and $\Psi_2\equiv(y\in\{b\}\lor z\in\{c\})$.
	Then both $\Psi_1$ and $\Psi_2$ are from $\C_2[U]$, but their disjunction is not.
	In fact, the least upper bound in $\C_2[U]$ for
	\[
		(x\in\{a\})\lor(y\in\{b\})\lor(z\in\{c\})
	\]
	is $\top$.
	\qed
\end{example}

\noindent

\subsection{\emph{Approximating 2-disjunctive Conjunctions}}\label{ss:dist_appox}

\emph{Any} CNF $\Psi$ over some set $Y$ of variables of bounded size can, in polynomial time,
be transformed into a DNF $\Psi'$.
Each DNF over two distinct variables $x,y$ can be brought into the canonical normal form

\begin{equation}
	\bigvee_{(a,b)\in L}	(x = a) \wedge (y = b)
	\label{def:normal_dnf}
\end{equation}
for some $L\subseteq U\times U$.
Conjunction and disjunction of two such normal forms then correspond to intersection and union of the respective subsets of
$U\times U$.

For arbitrary sets $Y$ of variables, though, it is non-trivial even to decide whether a given conjunction is different from $\bot$.

\begin{theorem}\label{t:const-NP}
	To decide for a formula $\Psi\in\C_2[U]$ whether or not $\Psi$ is satisfiable, i.e., different from $\bot$,
	is NP-complete.
\end{theorem}
\begin{proof}
	Since a satisfying assignment for $\Psi$ can be guessed and then checked in polynomial time,
	satisfiablity of $\Psi$ is in NP.
	NP-hardness, on the other hand, follows by a reduction from 3-colorability of graphs \cite{Beckert2000}.
	We illustrate the reduction with an example.
	\begin{example}\label{e:color}
For $\Vars=\{x_1,x_2,x_3,x_4\}$, consider the formula $\Psi$
\[
	\bigwedge_{\{x_i,x_j\}\in E}\begin{array}[t]{ll}
		\left(x_i\in\{b,c\}\vee x_j\in\{b,c\}\right)	&\wedge	\\
		\left(x_i\in\{a,c\}\vee x_j\in\{a,c\}\right)	&\wedge	\\
		\left(x_i\in\{a,b\}\vee x_j\in\{a,b\}\right)
	\end{array}
\]
where $E$ is given by
	\[
	E=\left\{ 	\{x_1,x_2\},
			\{x_1,x_4\},
			\{x_2,x_3\},
			\{x_3,x_4\},
			\{x_1,x_3\}	\right\}
	\]
Then $\Psi$ is satisfiable iff the undirected graph $(\Vars,E)$ has a 3-coloring.
In the given example, the graph
\begin{center}
\begin{tikzpicture}[
  node distance=0.7cm,
  every node/.style={circle, draw, inner sep=1pt}
]
\node (x1) {$x_1$};
\node (x2) [right=of x1] {$x_2$};
\node (x3) [below=of x1] {$x_3$};
\node (x4) [right=of x3] {$x_4$};
\draw (x1) -- (x2);
\draw (x1) -- (x3);
\draw (x1) -- (x4);
\draw (x2) -- (x3);
\draw (x2) -- (x4);
\draw (x3) -- (x4);
\end{tikzpicture}
\end{center}
cannot be colored by three colors.
Therefore, $\Psi$ is equivalent to $\bot$.
\qed
\end{example}
\end{proof}

\noindent
\emph{Exact} normalization (as defined in \cref{s:domain})
of a relation represented by some 2-CNF thus, in general, may be difficult to compute.
Instead of giving dedicated further abstraction techniques, we prefer to provide
for an \emph{arbitrary} relational domain $\R$,
a \emph{general} construction to approximate the 2-decomposable domain $\R_2$ further by a
2-decomposable domain $\R_2^\sharp$.
This construction is based on \emph{approximate} normalization.

Assume that an element in $\R_2$ is given by the meet $\bigsqcap R$
where $R$ is the collection $\angl{s_p}_{p\in\twoClusters}$ with $s_p\in\R^p$ ($p\in\twoClusters$).
According to \cref{t:2-decomp}, $\restr{(\bigsqcap R)}{p}\sqsubseteq s_p$ for all $p\in\twoClusters$.
As we have seen for 2-disjunctive constants, however, \emph{exact} normalization of $\bigsqcap R$, i.e.,
the values $\restr{(\bigsqcap R)}{p}$ may be hard to compute \emph{precisely}.
For an approximate normalization, we introduce a constraint system
in unknowns $r_p, p\in\twoClusters$ with the constraints
\begin{equation}
												\begin{array}{lll@{\;\;}r}
		r_{\{x,y\}}	&\sqsubseteq&	s_{\{x,y\}} &(x,y\in\X)		\\
		r_{\{x,y\}}	&\sqsubseteq&	\restr{(r_{\{x,z\}}\sqcap r_{\{z,y\}})}{\{x,y\}}
								&(x,y,z\in\X)
	\end{array}
	\label{def:constraints}
\end{equation}
This constraint system has already been considered for the normalization of $2$-projective domains~\cite{Schwarz23b}.
As all right-hand sides are monotonic,
the constraint system has a greatest solution -- whenever
each $\R^p,p\in\twoClusters,$ is a complete lattice.

In case that there is a greatest solution $\angl{r_p}_{p\in\twoClusters}$, $\restr{(\bigsqcap R)}{p}\sqsubseteq r_p$ holds
for all $p$, since $\angl{\restr{(\bigsqcap R)}{p}}_{p\in\twoClusters}$ is also a solution of the system \eqref{def:constraints}.
Then we call the collection $\angl{r_p}_{p\in\twoClusters}$ the \emph{approximate} normal form of the
collection $R$.
Here, we are not only interested in the \emph{existence} of a greatest solution of \eqref{def:constraints}
but also that it can be effectively computed.
For that, we consider the sets of
values possibly occurring during some fixpoint iteration for a particular
collection $R=\angl{s_p}_{p\in\twoClusters}$.

Let $I_\R[R]^p,p\in\twoClusters,$ be the least collection of sets such that
\begin{itemize}
\item	$s_p\in I_\R[R]^p$;
\item	If $r,r'\in\R_R^p$ then also $r\sqcap r'\in\R_R^p$;
\item	If $r\in I_\R[R]^{\{x,z\}}$ and $r'\in I_\R[R]^{\{z,y\}}$, then \\
	$\restr{(r\sqcap r')}{\{x,y\}}\in I_\R[R]^{\{x,y\}}$ for all $x,y,z\in\X$.
\end{itemize}
The sets $I_\R[R]^p$ collect the potential iterates occurring during greatest fixpoint iteration of \eqref{def:constraints}.
By construction, each set $I_\R[R]^p$ has a greatest element, namely, $s_p$, and is closed under binary $\sqcap$.
For the termination of Kleene fixpoint iteration for \eqref{def:constraints},
it suffices for each set $I_\R[R]^p$ to have a \emph{least}
element -- whose collection then coincides with the greatest solution of \eqref{def:constraints}.
This observation is summarized in the following proposition.

\begin{proposition}\label{p:fix}
	The following two statements are equivalent:
	\begin{enumerate}
	\item	For each $p\in\twoClusters$, $I_\R[R]^p$ has a least element;
	\item	The constraint system \eqref{def:constraints} has a greatest solution which
		can be attained by Kleene fixpoint iteration.
	\end{enumerate}
\end{proposition}

\begin{proof}
Assume that for each $p\in\twoClusters$, there is a least element $d_p\in I_\R[R]^p$.
We claim that $\underline R = \angl{d_p}_{p\in\twoClusters}$ is the greatest solution of \eqref{def:constraints}.
Since for each $p\in\twoClusters$, $d_p$ is a lower bound to all elements in $I_\R[R]^p$,
all constraints of \eqref{def:constraints} are satisfied. Therefore, $\underline R$ is a solution.
By induction on the definition of the sets $I_\R[R]^p$, any other solution $R'=\angl{r'_p}_{\twoClusters}$ consists of
lower bounds of these sets, i.e., $r'_p\sqsubseteq \bigsqcap I_\R[R]^p = d_p$ -- implying our claim.
To conclude statement (2),
it remains to prove that the greatest solution $\underline R$ can be reached by Kleene iteration.
For every $p$, $d_p$ is an element of the set $I_\R[R]^p$, and therefore, has arrived there after
finitely many applications of the inductive rule of their definitions.
Let $h$ be an upper bound to these numbers for all $d_p,p\in\twoClusters$.
Then, Kleene iteration for the constraint system \eqref{def:constraints} will also reach these values after at most
$h$ iterations.

For the reverse direction, assume that Kleene iteration for the greatest solution of \eqref{def:constraints}
terminates after $h$ iterations with a collection $\underline R= \angl{d_p}_{p\in\twoClusters}$.
By induction on the number $j$ of rounds, we  find each value $d^{(j)}_p$ attained for $r_p$, $p\in\twoClusters$,
after $j$ rounds, is an element of $I_\R[R]^p$. Therefore, $d_p= d_p^{(h)}\in I_\R[R]^p$ for all $p$.
It remains to prove that $d_p$ is also a lower bound of $I_\R[R]^p$.
To show this, we again proceed by induction, this time on the number $i$ of applications of the inductive rule
for the construction of the $I_\R[R]^p$, and prove that for all $i$ and any value $d'$ added to some set $I_\R[R]^p$ in the $i$th step,
it holds that $d_p^{(i)}\sqsubseteq d'$.
Therefore, $d_p$ is a lower bound to $I_\R[R]^p$ for all $p$, and statement (1) follows.
\qed
\end{proof}

\noindent
If all operations on abstract relations $r\in\R^{Y}$  for clusters $Y$ of size at most 3
are constant time and the height of all $\R[R]^p$ are bounded by $h$,
then the greatest solution of the constraint system \eqref{def:constraints}
can be computed in time polynomial in $h$ and the number of variables.

We call a relational domain 2-\emph{nice}, if the statements of \cref{p:fix} are satisfied for each
collection $R=\angl{s_p}_{p\in\twoClusters}$ with $s_p\in\R^p$.

Let us instantiate this construction to 2-disjunctive constants.
First, we note that the relational domain $\C[U]$ is finite and thus, in particular, 2-nice.
Let $\Psi=\angl{s_p}_{p\in\twoClusters}$ denote a collection with $s_p\in\C[U]^p$ for all $p$.
Assume that $\X$ consists of $n$ variables, and let $m$ be the number of constants occurring in any of the $s_p$.
According to the normal form \eqref{def:normal_dnf}, the lattice $I_{\C[U]}[\Psi]^p$
has height at most $m$ if $p$ consists of a single variable, and height bounded by $m^2$ if $p$ is a two-element set.
Since there are $\frac{1}{2}n(n+1)$ clusters, fixpoint iteration will terminate after
${\cal O}(n^2\cdot m^2)$ updates.
\qed

\noindent
Due to NP-hardness of satisfiability, we cannot expect
the greatest solution of the constraint system for 2-disjunctive constants to always return the exact normal form.
For the formula from \cref{e:color}, e.g., it
returns for each pair $\{x_i,x_j\}\in E$, $i\neq j$,
\[
	\begin{array}{l}
	\left(x_i=a\wedge x_j\in\{b,c\}\right)\vee
	\left(x_i=b\wedge x_j\in\{a,c\}\right)\vee \\
	\quad \left(x_i=c\wedge x_j\in\{a,b\}\right)
	\end{array}
\]
-- which is different from $\bot$.

For a relational domain $\R$,
we call a collection $R = \angl{r_p}_{p\in\twoClusters}$
with $r_p\in\R^p$ for all $p$, \emph{stable}
if it is a solution of the constraint system \eqref{def:constraints} with $s_p \equiv r_p$.
We remark that stability of $R$ implies that, if $r_p=\bot$ for some $p$, then $r_{p'}=\bot$
for all other $p'\in\twoClusters$ as well.
Now we introduce for a relational domain $\R$
the domain $\R_2^\sharp$ of all
stable collections.
The ordering $\sqsubseteq^\sharp$ on the domain $\R_2^\sharp$ is defined by $R\sqsubseteq^\sharp R'$ if
$r_p\sqsubseteq r'_p$ for all $p\in\twoClusters$ when
$R = \angl{r_p}_{p\in\twoClusters}$ and
$R' = \angl{r'_p}_{p\in\twoClusters}$.
Thus, $(\bigsqcap R)\sqsubseteq(\bigsqcap R')$ whenever $R\sqsubseteq^\sharp R'$.

Abstract join as well as abstract restriction for $\R_2^\sharp$ then is modeled along the definitions of
join and restriction for $\R_2$,
but refers to the representation as solution to the constraint system \eqref{def:constraints}.
For $R=\angl{r_p}_{p\in\twoClusters}$, $R'=\angl{r'_p}_{p\in\twoClusters}$ in $\R_2^\sharp$, we define the abstract join by
\[
	R\sqcup^\sharp R'= \angl{r_p\sqcup r'_p}_{p\in\twoClusters}
\]
while for $Y\subseteq\X$, and $R=\angl{r_p}_{p\in\twoClusters}$, we define abstract restriction by
\[
	\begin{array}{lll}
		\restrs{\angl{r_p}_{p\in\twoClusters}}{Y}
			&=&\angl{\restr{r_p}{Y}}_{p\in\twoClusters}	\\
			&=&\angl{\restr{r_p}{Y\cap p}}_{p\in\twoClusters}	\\
	\end{array}
\]
where the latter equality follows since for $r_p\in\R^p$, $\restr{r_p}{p} = r_p$.
We have:

\begin{proposition}\label{p:abstract}
Assume that $\R$ is 2-nice and satisfies \eqref{def:decomp2}. Then we have:
\begin{enumerate}
\item
For each $R,R'\in\R_2^\sharp$, also $R\sqcup^\sharp R'$ is again in $\R_2^\sharp$ and
is the least upper bound of $R,R'$. Moreover,
\[
(\bigsqcap R)\sqcup(\bigsqcap R')\sqsubseteq\bigsqcap (R\sqcup^\sharp R')
\]
\item
For each $R\in\R_2^\sharp$ and $Y\subseteq\X$,
$\restrs{R}{Y}$ is again in $\R_2^\sharp$ where
\[
\restr{(\bigsqcap R)}{Y}\sqsubseteq \bigsqcap(\restrs{R}{Y})
\]
holds.
\item
For each $R=\angl{r_p}_{p\in\twoClusters}$ ,$R'=\angl{r'_p}_{p\in\twoClusters}$ in $\R_2^\sharp$,
the greatest lower bound $R\sqcap^\sharp R' = \angl{r''_p}_{p\in\twoClusters}$ is determined as the greatest solution
of \eqref{def:constraints} with start values
$s_p = r_p\sqcap r'_p$ ($p\in\twoClusters$).
\end{enumerate}
\end{proposition}

\begin{proof}
For the first statement, let $R=\angl{r_p}_{p\in\twoClusters}$ and $R'=\angl{r'_p}_{p\in\twoClusters}$.
As the ordering on $\R_2^\sharp$ is componentwise, it suffices to prove that $R\sqcup^\sharp R'$ is again in $\R_2^\sharp$,
i.e., the collection $r_p\sqcup r'_p,p\in\twoClusters,$ is a solution of the constraints in \eqref{def:constraints}.
For this, we calculate:
\[
\begin{array}{lll}
\multicolumn{3}{l}{r_{\{x,y\}}\sqcup r'_{\{x,y\}}}	\\
\qquad\qquad&\sqsubseteq&
\restr{(r_{\{x,z\}}\sqcap r_{\{z,y\}})}{\{x,y\}}\sqcup
\restr{(r'_{\{x,z\}}\sqcap r'_{\{z,y\}})}{\{x,y\}}		\\
&\sqsubseteq&
\restr{((r_{\{x,z\}}\sqcup r'_{\{x,z\}})\sqcap (r_{\{z,y\}}\sqcup r'_{\{z,y\}}))}{\{x,y\}}	\\
\end{array}
\]
for all variables $x,y,z\in\X$. From that, the statement follows.

To prove the second statement, we must verify that the collection $\restr{r_p}{Y\cap p}, p\in\twoClusters$
satisfies all constraints in \eqref{def:constraints}. Indeed, we find by monotonicity,
\[
\begin{array}{lll}
\restr{ r_{\{x,y\}}} {Y}
&\sqsubseteq&
\restr{ (r_{\{x,z\}}\sqcap r_{\{z,y\}})} {\{x,y\}\cap Y}	\\
&\sqsubseteq&
\restr{ (\restr{r_{\{x,z\}}}{Y}\sqcap \restr{r_{\{z,y\}}}{Y}) }{\{x,y\}\cap Y}	\\
\end{array}
\]
for all $x,y,z\in\X$, and the claim follows.
The final statement then follows from the definition.
\qed
\end{proof}

\noindent
Elements of $\R_2^\sharp$ are collections $\angl{r_p}_{p\in\twoClusters}$. For every $p\in\twoClusters$, we can
consider elements $r_p\in\R^p$ as elements of $\R_2^\sharp$ as well by assuming that $r_p$ represents the stable collection
$\angl{\restr{r_p}{q}}_{q\in\twoClusters}$.

According to \cref{p:abstract}, both joins and restrictions can be computed componentwise.
As a consequence, we find:

\begin{theorem}\label{t:abstract}
For a 2-nice relational domain $\R$ which satisfies \eqref{def:decomp2},
the domain $\R_2^\sharp$ is a 2-decomposable relational domain.
\qed
\end{theorem}

\begin{figure}[h]
\begin{center}
\scalebox{0.8}{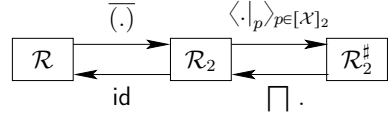}
\end{center}
\caption{\label{f:rel}The relationship between abstract relational domains.}
\end{figure}

\noindent
\cref{f:rel} shows the abstract relational domains $\R,\R_2$, and $\R_2^\sharp$ together with the mappings between them.
\noindent
According to \cref{t:abstract},
the domain $\C_2^\sharp[U]$ of abstract 2-disjunctive constants
is indeed 2-decomposable.
The given construction provides us with polynomial algorithms for
least upper bound, greatest lower bound, and projection.

\subsection{\emph{Assignments}}\label{ss:const-ass}
Let us return to the relational domain $\C_2[U]$ of 2-dis\-junc\-tive constants and indicate how
abstract transformers for assignments $x\,{:=}\,s$ can be tailored.
For 2-dis\-junc\-tive constants,
we only consider right-hand sides $s$ where $s$ is either $?$ (unknown value),
or of the form $A |y_1|\ldots|y_k$ where $A$ is a set of constants and $y_1,\ldots,y_k\in\X$ are variables.
The concrete semantics of such an assignment is given by
\[
	\begin{array}{lll}
		\sem{x\,{:=}\,?}\,\Sigma 	&=&
		\{\sigma\oplus\{x\mapsto c\}\mid \sigma\in \Sigma, c\in U\}	\\
		\sem{x\,{:=}\,A| y_1|\ldots| y_k}\,\Sigma 	&=&
		\{\sigma\oplus\{x\mapsto a\}\mid \sigma\in \Sigma, a\in A\} \cup\\
			& &
		\bigcup_{j=1}^k
		\{\sigma\oplus\{x\mapsto \sigma\,y_j\}\mid \sigma\in \Sigma\}
	\end{array}
\]
Generalizing the corresponding abstract semantics for (copy) constant propagation, we define the logic transformer
for $\C_2[U]$ by
\[
	\begin{array}{lll}
\sem{x\,{:=}\;?}_2\,\Psi&=&	\restr{\Psi}{\X\setminus\{x\}}	\\
\sem{x\,{:=}\;A| y_1|\ldots| y_k}_2\,\Psi&=&
			(x\in A)\land\restr{\Psi}{\X\setminus\{x\}}\sqcup_2	\\
			&&
			\bigsqcup_{2\;j=1}^{\phantom{2\;}k}\;\;\sem{x\;{:=}\,y_j}_2 \,\Psi
	\end{array}
\]
\begin{proposition}\label{s:const_ass}
	\begin{enumerate}
		\item	The logic transformer $\sem{x\,{:=}\,?}_2$ is precise, i.e.,
			\begin{equation}
				\sem{x\,{:=}\,?}\,(\gamma\,\Psi) = \gamma\,(\sem{x\,{:=}\,?}_2\,\Psi)
				\label{def:precise_restr}
			\end{equation}
			In particular, it is distributive and commutes with $\bot$.
		\item	The logic transformer $\sem{x\,{:=}\,A\mid y_1|\ldots| y_k}_2$ is precise,
			if  the logic transformers for $x\,{:=}\,y_j$, $j=1,\ldots,k$, are.
	\end{enumerate}
\end{proposition}
Thus, we have reduced the construction of logic transformers for assignments to restriction and the
construction of logic transformers for variable-variable
assignments $x\,{:=}\,y$. For $y\equiv x$, the assignment is the identity, i.e., we set
$\sem{x\,{:=}\,x}_2\,\Psi = \Psi$.
Therefore, assume that $y$ is different from $x$, and assume that
$\restr{\Psi}{\X\setminus\{x\}} = \Psi'$.
Let $B$ denote the set of constants so that $\restr{\Psi'}{\{y\}}$ equals $y\in B$.
Let $\Psi_y$ denote the conjunction of all formulas $\restr{\Psi'}{p}$ for $p\in\twoClusters$ with $y\in p$.
Let $\Psi'' = \Psi_y[x/y]$ denote the formula obtained from $\Psi_y$ by renaming each occurrence of the variable $y$ with $x$.
Then we define
\[
	\sem{x\,{:=}\,y}_2\,\Psi = \Psi'\wedge\left(\bigvee_{a\in B}x=a\wedge y=a\right)\wedge\Psi''
\]
Let $\bar\Psi$ denote the formula returned by that transformer for $\Psi$.
Intuitively, our definition means for $x\not\in p$, that
$\restr{\bar\Psi}{p}=\restr{\Psi}{p}$, i.e., $\restr{\Psi}{p}$ is preserved while additionally,
$\restr{\bar\Psi}{\{x\}} = \restr{\Psi}{\{y\}}[x/y]$,
$\restr{\bar\Psi}{\{x,y\}} = \bigvee_{a\in B} x=b\wedge y=b$,
and for $z\not\in\{x,y\}$,
$\restr{\bar\Psi}{\{x,z\}} = \restr{\Psi}{\{y,z\}}[x/y]$.

\begin{proposition}
	The logic transformer $\sem{x\,{:=}\,y}_2$ is precise, i.e.,
	\begin{equation}
		\sem{x\,{:=}\,y}\,(\gamma\,\Psi) = \gamma\,(\sem{x\,{:=}\,y}_2\,\Psi)
		\label{def:precise_var}
	\end{equation}
	holds.
					\qed
\end{proposition}

\noindent
The same construction allows us to construct \emph{abstract} logic transformers
$\sem{x\,{:=}\,s}_2^\sharp:C_2^\sharp[U]\to C_2^\sharp[U]$ --
only that the least upper bound operation and projection of $\C_2[U]$ must be replaced by the
corresponding operations of $\C_2^\sharp[U]$.
The abstract transformer then, however, is only \emph{sound} and no longer \emph{precise},
since the projection operation of $\C_2^\sharp[U]$ may return for an abstract relation $R$ whose concretization is \emph{empty}
an abstract relation with a \emph{non-empty} concretization.
Accordingly, \cref{def:precise_restr} and \cref{def:precise_var} may be violated.

\subsection{\emph{Guards}}\label{ss:guards}

It remains to provide the semantics of guards. Again, we first consider
the domain $\C_2[U]$ of 2-disjunctive formulas (modulo logical equivalence),
ordered by implication.
We consider positive guards of the form $x\in A$, and conversely,
negative guards of the form $x\not\in A$.
Positive guards thus can directly be expressed in $\C_2[\U]$.
Thus we set
\begin{equation}
\sem{?(x\in A)}\,\Psi = \Psi \wedge (x\in A)
\label{def:const_guard}
\end{equation}
Negative guards on the other hand cannot be directly expressed in $\C_2[U]$ -- at least
if there are unknown constant values beyond the finite universe $U$. To deal with this,
we introduce a dedicated
fresh symbol $\bullet\not\in U$ with the understanding that
$\bullet$ repesents any value $a\not\in U$.
The property $x\not\in A$ then can equivalently be represented by
\[
x\in(U\cup\{\bullet\})\setminus A
\]
allowing us to deal with such co-finite sets of possible values in the same way as we
did for finite sets of values alone.

\newcommand{\impliess}{\,\longrightarrow^\sharp\,}
\section{Directed Relational Domains}\label{s:directed}

Instead of plain equalities, let us now consider \emph{in}equalities between variables and constants instead of equalities and
abandon disjunctions. We will, however, add disjunctions in the end as well.
Thus for now, we just consider finite conjunctions of inequalities
of the form
\[
	d\sqsubseteq x,\quad x\sqsubseteq y,\quad
	\text{or}\quad x\sqsubseteq d
\]
for variables $x,y\in\X$ and constant values $d$.
As usual, we consider conjunctions only up to semantic equivalence.
We call inequalities of the form $d\sqsubseteq x$ lower bound constraints, and $d$ a lower bound for $x$.
Analogously for upper bounds. Inequalities of the form $x\sqsubseteq y$ are called variable constraints.

Assume we are given a partial order (po), i.e., a set $P$ partially ordered by some relation $\leq$.
Examples of partial orders of interest are
\begin{description}
	\item[\textsf{Subsets.}]
		The set $2^U$ of all subsets of some finite universe $U$ where the ordering is
		subset inclusion $\subseteq$;
	\item[\textsf{Integers.}]
		The set $\mathbb{Z}$ of integers equipped with the natural ordering $\leq_\Z$;
	\item[\textsf{Multisets.}]
		Multisets, i.e., the set of all mappings $\mu:U\to\mathbb{N}$ from
		elements in $U$ to their multiplicities ordered by multiset inclusion
		$\subseteq_\N$.
	\item[\textsf{Strings.}]
		The set of all strings $\Sigma^*$ for some finite alphabet $\Sigma$.
		Several partial orderings are of interest:
		\begin{itemize}
			\item	the \emph{prefix} ordering
				$\leq_p$; e.g., $\textsf{ab} \leq_p \textsf{abcd}$;
			\item	the \emph{substring} ordering $\leq_s$,
				e.g., $\textsf{bc}\leq_s \textsf{abcde}$;
			\item	the \emph{scattered substring} ordering $\leq_{ss}$,
				e.g., $\textsf{bd}\leq_{ss} \textsf{abcde}$.
		\end{itemize}
\end{description}
Much more expressive constraints on strings have been studied, e.g., in
\cite{ChenCHLW18,DayGGM23,AbdullaADHJ19,ganesh2011decidable}.
In particular, for a fragment containing the prefix ordering, decision procedures are known based on
(synchronous) multi-tape finite automata \cite{YuBH11}.
Due to their expressiveness, these techniques come with a considerable computational effort.
Instead, we follow \citet{Arceri22} where basic relational domains are considered for reasoning
about variables of string type, sets (of characters), or integers (lengths of strings).
Their analyses relate program variables only according to some
partial order, and also consider lower bounds.
Here, these considerations are complemented by taking \emph{upper bounds} into account as well and,
eventually, by adding disjunctions.

A mapping $\sigma:\X\to P$ is a model of $\Psi$ (relative to $ P$), written as $\sigma\models\Psi$, if
$\Psi\neq\bot$, and
\begin{itemize}
	\item $d\leq\sigma\,x$ (in $ P$) for each constraint $d\sqsubseteq x$ in $\Psi$;
	\item $\sigma\,x\leq d$ (in $ P$) for each constraint $x\sqsubseteq d$ in $\Psi$; and
	\item $\sigma\,x\leq\sigma\,y$ (in $ P$) for each constraint $x\sqsubseteq y$ in $\Psi$.
\end{itemize}
Let $\D[P]$ denote all finite conjunctions over $ P$ modulo semantic equivalence
where the ordering on $\D[P]$ is semantic implication.
As before, normal forms of conjunctions will be considered up to reordering of atomic propositions.
Thus, syntactic equality of conjunctions here means equality of the respective \emph{sets} of propositions.
Let $\Psi$ denote a finite conjunction where $V\subseteq P$ is the set of values occurring in $\Psi$ as lower or upper bounds.
To provide a first normal form for $\Psi$, we proceed in two steps.
First, we determine the transitive closure $(\leq\cup\sqsubseteq)^+$
on the set $\X\cup V$ of the constraints provided by $\Psi$.
In case that $(a,b)\in(\leq\cup\sqsubseteq)^+$ for $a,b\in V$
where $a\leq b$ does not hold in $ P$, then $\Psi$ is unsatisfiable
and therefore represented by the dedicated element $\Psi'=\bot$.
If this is not the case, let $\Psi'$ denote the conjunction of all inequalities
$s_1\sqsubseteq s_2$ where $(s_1,s_2)\in(\leq\cup\sqsubseteq)^+$
and either $s_1$ or $s_2$ or both are in $\X$.

In the second step, when $\Psi'\neq\bot$,
we remove all redundant constraints.
These are constraints of the form
\begin{itemize}
\item	$x\sqsubseteq x$ for $x\in\X$, as these constraints hold vacuously;
\item	$a\sqsubseteq x$ for $a\in V$ and $x\in\X$ if there is also a constraint $b\sqsubseteq x$ with $a\leq b$, i.e., there is a stricter lower bound;
\item	$x\sqsubseteq b$ for $b\in V$ and $x\in\X$ if there is also a constraint $x\sqsubseteq a$ with $a\leq b$, i.e., there is a stricter upper bound.
\end{itemize}
Additionally, we set $\Psi'$ to $\bot$ whenever for some variable $x$,
\begin{itemize}
\item		there is no lower bound in $ P$ for the set of upper bounds
	provided for $x$ by $\Psi$; or
\item		there is no upper bound in $ P$ for the set of lower bounds
	provided for $x$ by $\Psi$.
\end{itemize}
Assume, e.g., that $\Psi$ is given by
\[
(\textsf{abc}\sqsubseteq x)\wedge(\textsf{abd}\sqsubseteq x)
\]
where we consider the prefix order $\leq_p$ on strings.
Since $\textsf{abc},\textsf{abd}$ cannot be prefixes of the same string, this conjunction
is considered equivalent to $\bot$.

Let us denote the resulting conjunction $\Psi'$ by $\textsf{nf}_0[\Psi]$
and call it the 0-normal form of $\Psi$.
Assuming that comparisons of values
as well as checks for common lower or upper bounds
are constant-time operations,
0-normal forms can be computed in polynomial time.

\subsection{\emph{Lattice Domains}}\label{ss:lattice}

An important special case is when $ P$ is a \emph{lattice}, i.e., a po
where every two elements $a,b$ both have
a least upper bound $a\vee b$ and a greatest lower bound $a\wedge b$.

\begin{example}\label{e:lattice}
The po $2^U$ ordered by subset inclusion is a complete lattice and thus, in particular, a lattice.
The integers $\mathbb{Z}$ with the natural ordering is another example of a
lattice, this time without \emph{least} or \emph{greatest} element.
Yet another example are \emph{multisets}: this lattice has a \emph{least}, but no \emph{greatest} element.

The po $\Sigma^*$ of strings ordered by the prefix relation is not a lattice.
$\Sigma^*$ provides a least element $\epsilon$, as well as greatest lower bounds, namely,
the maximal common prefix,
but does not have least upper bounds to all pairs of strings.
There is, for example, no upper bound to $\textsf{abc}$ and $\textsf{abd}$ in $\Sigma^*$.
\qed
\end{example}

\noindent
When $ P$ is a lattice, we can provide a dedicated normal form which, however, may now use
constants from $ P$
which did not occur in $\Psi$ before.
Assume now that $\Psi'$ is the 0-normal form of $\Psi$.
If $ P$ has a least element $\bot_P$, we add the vacuous constraint $\bot_P\sqsubseteq x$ to every variable $x$.
Likewise, if $P$ has a greatest element $\top_P$, we add the constraint $x\sqsubseteq\top_P$.

If $\Psi'$ is different from $\bot$, we subsequently simplify $\Psi'$ further
by replacing for each variable $x\in\X$,
\begin{itemize}
\item	the set of upper bound constraints occurring in $\Psi'$, if it
	is non-empty and consists of $(x\sqsubseteq b_1)\land\ldots\land(x\sqsubseteq b_r)$,
	with the single constraint $(x\sqsubseteq(\bigwedge_{i=1}^r b_i))$;
\item	the set of lower bound constraints in $\Psi'$, if it
        is non-empty and consists of $(a_1\sqsubseteq x)\land\ldots\land(a_r\sqsubseteq x)$,
	with the single constraint $((\bigvee_{i=1}^r a_i)\sqsubseteq x)$.
\end{itemize}
Let us denote the resulting formula by $\textsf{nf}_1[\Psi]$ and call it the 1-normal form of $\Psi$.
The 1-normal form of $\Psi$ can be computed in polynomial time as well -- given that comparisons as well as
pairwise least upper bounds and greatest lower bounds in $P$ are constant time.
We have:
\begin{theorem}\label{t:lattice_directed}
	Assume that the po $P$ is a lattice.
	Then the following holds:
	\begin{enumerate}
		\item 	A conjunction $\Psi$ is satisfiable over $P$ iff $\textsf{nf}_1[\Psi]\neq\bot$.
		\item	For arbitrary conjunctions $\Psi_1,\Psi_2$ over $P$,
				$\Psi_1\implies\Psi_2$ iff $\textsf{nf}_1[\Psi_1] = \textsf{nf}_1[\Psi_1\land\Psi_2]$.
	\end{enumerate}
	Satisfiability as well as implication are decidable in polynomial time.
	\qed
\end{theorem}

\begin{proof}
If $\Psi'=\textsf{nf}_1[\Psi] = \bot$, then $\Psi$ cannot be satisfiable since any of the simplification steps preserves
the set of satisfying assignments.
So, assume that $\Psi'$ is syntactically different from $\bot$.
Let $\sigma$ be the variable assignment which maps each variable $x$
to its lower bound $a_x\in P$ -- if it exists, and to some fixed element $\underline a$ which is less or equal to any
other lower bound mentioned in $\Psi'$.
Then all single variable constraints are satisfied as well as, by transitivity,
all constraints $x\sqsubseteq y$ occurring in $\Psi'$. Therefore, $\sigma\models\Psi$ -- implying that $\Psi$ is satisfiable.
From this, statement (1) follows.

To prove statement (2), consider conjunctions $\Psi'_1,\Psi'_2$ both in 1-normal form.
If these syntactically coincide, then obviously also
$\Psi'_1 \iff \Psi'_2$ holds.
For the reverse direction, we prove that if $\Psi'_i$ are distinct, then they cannot be equivalent.
From that, the assertion follows.
If one of them equals $\bot$ and the other not, then by statement (1), they cannot be equivalent.
Therefore, assume that both are satisfiable and thus, different from $\bot$.
We consider all cases how the $\Psi_i$ may differ.

\begin{description}
\item[\textsf{Lower bounds.}]
First, assume that there are constraints $a_i\sqsubseteq x$, $i=1,2$, for some variable $x$ in $\Psi'_i$
where $a_1$ is different from $a_2$.
Assume w.l.o.g.\ that $a_1\not\leq a_2$ holds. Let
$L_x$ denote the set consisting of $x$ together with  variables $z\in\X$ where $\Psi'_2$ has a constraint $z\sqsubseteq x$.
Let $\sigma$ denote some assignment with $\sigma\models\Psi'_2$.
Then we construct a variable assignment $\sigma'$ such that $\sigma' \models \Psi'_2$ but $\sigma' \not\models \Psi'_1$ by

\[
	\sigma'\,z = \begin{cases}
		\sigma\,z\wedge a_2 & \text{if }z\in L_x\\
		\sigma\,z & \text{otherwise}
	\end{cases}
\]
Then still $\sigma'\models\Psi'_2$.
But since $a_1\not\leq a_2$, it follows that $\sigma'$ does not satisfy $a_1 \sqsubseteq x$ and thus it does not model $\Psi'_1$.

If there is a constraint $a_1\sqsubseteq x$ in $\Psi'_1$, but no lower bound constraint for $x$ in $\Psi'_2$,
then there is some value $\underline\bot\in P$ different from $a_1$ so that $\underline\bot\leq a_1\wedge\sigma\,x$ holds.
This value allows us to construct an analogous distinguishing assignment $\sigma'$ where we use $\underline\bot$
instead of $a_2$.

\item[\textsf{Upper bounds.}]
First, assume that there are constraints $x\sqsubseteq b_i$, $i=1,2$, for some variable $x$ in $\Psi'_i$
where $b_1$ is different from $b_2$.
W.l.o.g., assume that $b_2\not\leq b_1$.
Let $U_x\subseteq\X$ denote the subset consisting of $x$ together with all unknowns $z$ where $\Psi'_2$
has a constraint $x\sqsubseteq z$.
Let $\sigma$ denote some assignment with $\sigma\models\Psi'_2$.
Then we construct a variable assignment $\sigma'$ by:
\[
	\sigma'\,z = \begin{cases}
		\sigma\,z\vee b_2 & \text{if }z\in U_x\\
		\sigma\,z & \text{otherwise}
	\end{cases}
\]
Then
still $\sigma'\models\Psi'_2$ holds.
But since $b_2\not\leq b_1$, $\sigma'$ does not satisfy $\Psi'_1$.

If there is a constraint $x\sqsubseteq b_1$ in $\Psi'_1$, but no upper bound constraint for $x$ in $\Psi'_2$,
we introduce a value $\overline\top\in P$ which is different from $b_1$ with $(b_1\vee\sigma\,x)\leq \overline\top$,
and construct an analogous distinguishing assignment $\sigma'$ only that we use $\overline\top$ instead of $b_2$.

\item[\textsf{Variable  Constraints.}]
Assume that, w.l.o.g., $\Psi'_1$ has a constraint $(x\sqsubseteq y)$ for $x,y\in\X$ which does not occur in $\Psi'_2$
where we assume that for every variable $z\in\X$ both lower and upper bounds are provided by $\Psi'_1$ iff they are
provided by $\Psi'_2$ and that, whenever they are provided, they agree.
Consider again the set $U_x$ of $x$ together with all variables $z$ with constraints $x\sqsubseteq z$,
and the set $L_y$ of $y$ together with all variables $z$ with constraints $z\sqsubseteq y$
occurring in $\Psi'_2$. Since $x\sqsubseteq y$ does not occur in $\Psi'_2$, $U_x\cap L_y=\emptyset$.

Let $\sigma$ denote an assignment with $\sigma\models\Psi'_2$.
First assume that $\Psi'_2$ has constraints $x\sqsubseteq b$ and $a\sqsubseteq y$.
From $x \sqsubseteq y $ not occurring in $\Psi'_2$, it follows that $b \not\leq a$.
Now we construct an assignment $\sigma'$ by:
\[
\sigma'\,z= \begin{cases}
  b\vee \sigma\,z & \text{if } z\in U_x \cup \{x\}\\
  a\wedge \sigma\,z & \text{if } z\in L_y \cup \{y\}\\
  \sigma\,z & \text{otherwise}
  \end{cases}
\]
Then $\sigma'\models\Psi'_2$, while $\sigma'\,x=b$ and $\sigma'\,y=a$.
As $b \not\leq a$, $\sigma'$ does not fulfill the constraint $x \sqsubseteq y$ from $\Psi'_1$.

If no upper bound of $x$ is provided,
we choose some value $b$ strictly larger than $\sigma\,x\vee\sigma\,y$, and define a variable assignment $\sigma'$ by
$\sigma'\,z= b\vee \sigma\,z$ for $z\in U_x$,
and
$\sigma'\,z= \sigma\,z$ otherwise.
Then $\sigma'\models\Psi'_2$. In order to additionally satisfy $x\sqsubseteq y$, we would have
$\sigma'\,x = b\vee\sigma\,x = b\leq \sigma'\,y$ -- which is impossible.

Likewise, if no lower bound of $y$ is provided,
we choose some value $a$ strictly less than $\sigma\,x\wedge\sigma\,y$, and define a variable assignment $\sigma'$ by
$\sigma'\,z= a\wedge \sigma\,z$ for $z\in L_y$,
and $\sigma'\,z= \sigma\,z$ otherwise.
Then $\sigma'\models\Psi'_2$. In order to additionally satisfy $x\sqsubseteq y$, we would have
$\sigma'\,x = \sigma\,x \leq \sigma'\,y = a$ -- which again is impossible.
\end{description}
\qed
\end{proof}

\noindent
For lattices, therefore, the construction of normal forms
allows deciding satisfiability as well as semantic implication.
From our examples, sets, integers, and multisets are lattices.
Strings, ordered by the prefix relation, on the other hand, already do not form a lattice anymore.
This po, however, is \emph{bounded-complete}.
Recall that a po $ P$ is bounded-complete if every subset $A\subseteq P$ which has some upper bound,
also has a \emph{least} upper bound.
When $ P$ is bounded-complete, then we at least know that
\begin{itemize}
\item	every non-empty subset $B\subseteq P$ has a \emph{greatest} lower bound; and
\item	$ P$ has a \emph{least} element $\bot_P$.
\end{itemize}
Thus, every formula $\Psi$ over a bounded-complete po $P$ which provides some upper bound to every variable $x\in\X$
also can be brought into 1-normal form. Let us call such conjunctions \emph{bounded}. We obtain:

\begin{proposition}\label{p:lattice_directed}
          Given a po $P$ that is bounded-complete, the following holds:
          \begin{enumerate}
          \item   A bounded conjunction $\Psi$ is satisfiable over $P$ iff $\textsf{nf}_1[\Psi]\neq\bot$.
          \item   For arbitrary bounded conjunctions $\Psi_1,\Psi_2$ over $P$,
                  $\Psi_1\implies\Psi_2$ iff $\textsf{nf}_1[\Psi_1] = \textsf{nf}_1[\Psi_1\land\Psi_2]$. \qed
          \end{enumerate}
\end{proposition}

\noindent
When we drop the extra assumption that conjunctions are bounded, \cref{p:lattice_directed} need no longer hold.

\begin{example}\label{e:prefix}
	For prefixes of strings, consider the conjunction
	\[
	(\textsf{ab}\sqsubseteq x)\wedge(x\sqsubseteq \textsf{abc})\wedge (\textsf{abd}\sqsubseteq y)\wedge(x\sqsubseteq y)
	\]
				This formula is semantically equivalent to
	\[
	(\textsf{ab}\sqsubseteq x)\wedge(x\sqsubseteq \textsf{ab})\wedge (\textsf{abd}\sqsubseteq y)\wedge(x\sqsubseteq y)
	\]
	although the formulas are syntactically different.

	Even without upper bounds, not all implications can be inferred via transitive closure alone.
	Again for prefixes of strings, consider
	\[
	\begin{array}{c}
	(\textsf{abc}\sqsubseteq y_1)\wedge
	(\textsf{abd}\sqsubseteq y_2)\wedge
	(x\sqsubseteq y_1)\wedge(x\sqsubseteq y_2)\wedge
	(\textsf{ab}\sqsubseteq z)
	\end{array}
	\]
	The first four constraints imply that $x\sqsubseteq \textsc{ab}$, which, by the last constraint, implies that
	$x\sqsubseteq z$ must hold as well.
			\qed
\end{example}

\noindent
For a conjunction $\Psi$ and a subset $Y\subseteq\X$ of variables, let $\restrs{\Psi}{Y}$ yield $\bot$ if $\Psi$ equals $\bot$,
and otherwise, yield the conjunction of all constraints in $\Psi$ that only uses variables from $Y$.

For conjunctions $\Psi_1,\Psi_2$ in 1-normal form and different from $\bot$, we define the \emph{abstract join}
$\Psi_1\sqcup^\sharp\Psi_2$ as the conjunction of the following constraints:
\begin{itemize}
\item	all constraints $x\sqsubseteq y$, $x,y\in\X$, which occur both in $\Psi_1$ and $\Psi_2$;
\item	all constraints $(d_1\wedge d_2)\sqsubseteq x$, $d_1,d_2\in P$, $x\in\X$ where $d_i\sqsubseteq x$ occurs in $\Psi_i$;
\item	all constraints $x\sqsubseteq(d_1\vee d_2)$, $d_1,d_2\in P$, $x\in\X$ where $x\sqsubseteq d_i$ occurs in $\Psi_i$.
\end{itemize}
Then we have:

\begin{theorem}\label{t:lattice_2_dec}

Assume that $ P$ is a lattice.
\begin{enumerate}
\item	If $\Psi$ is a conjunction in 1-normal form, then for every subset $Y\subseteq\X$,
	$\restr{\Psi}{Y}$ is given by $\restrs{\Psi}{Y}$ where the latter conjunction
	is again in 1-normal form.
\item	For $\Psi_1,\Psi_2$ in 1-normal form, $\Psi_1\sqcup^\sharp \Psi_2$ is the least upper bound of $\Psi_1,\Psi_2$ in
	$\D[P]$.
\item	The domain $\D[P]$ is a 2-decomposable relational domain.	\qed
\end{enumerate}
\end{theorem}

\noindent
While statement (1) of \cref{t:lattice_2_dec} remains true also for bounded conjunctions over a bounded-complete po,
the least upper bound of two bounded conjunctions need no longer be bounded, as the least upper bounds of
the respective upper bounds need not exist.
For the prefix ordering on $\Sigma^*$, e.g., we have
\[
(x\sqsubseteq \textsf{abc})\sqcup(x\sqsubseteq \textsf{abd}) = \top
\]
i.e., all information about upper bounds is lost.

\subsection{\emph{The General Case}}\label{ss:general_directed}
For general (even finite) partial orders, the dedicated constructions for lattices cannot be directly applied.
Already the problem of determining whether or not a conjunction
is satisfiable, turns out to be surprisingly difficult.
Assume that elements in $P$ can be represented and compared in polynomial time.
Then we find:

\begin{theorem}\label{t:directed-complexity}
The problem of determining for a given partial order $P$ and a conjunction $\Psi$, whether $\Psi$
is satisfiable over $P$, is NP-complete.
\end{theorem}

\begin{proof}
				Since a satisfying assignment for a conjunction $\Psi$ can be guessed in polynomial time,
	it remains to prove the hardness part.
		For that, consider the problem of 3-colorability of an undirected finite graph $G = (V,E)$.
	Let $v_1,\ldots, v_n$ be an enumeration of the vertices in $V$.
		Then, we construct a partial order $P$ consisting of the elements
	\[
		\{\angl{v_i,c}\mid i=1,\ldots,n, c=1,2,3\}\;
		\begin{array}[t]{@{}l}
		\dot\cup\;\{\underline v_i\mid i=1,\ldots,n\}\;\\
		\dot\cup\;\{\overline v_i\mid i=1,\ldots,n\}
		\end{array}
	\]
	where the partial ordering $\leq$ of $P$ is the least partial order
	satisfying
	\[
		\begin{array}{lll@{\quad}l}
			\angl{v_i,c}&\leq&	\angl{v_j,c'}	&\text{whenever}\;
						\{v_i,v_j\}\in E\land i<j\land c\neq c'	\\
			\angl{v_i,c}&\leq&	\overline v_i	&\text{whenever}\;
						\exists\,j>i.\,\{i,j\}\in E\\
			\underline v_j	&\leq&	\angl{v_j,c}	&\text{whenever}\;
						\exists\,i<j.\,\{i,j\}\in E\\
		\end{array}
	\]
	For $P$, we define a conjunction $\Psi$ in the variables $x_i,i=1,\ldots,n$, by
	\[
		\begin{array}{l}
						\bigwedge_{\{v_i,v_j\}\in E,i<j}
			(x_i\sqsubseteq\overline v_i)
			\wedge (x_i\sqsubseteq x_j)
			\wedge(\underline v_j\sqsubseteq x_j)
		\end{array}
	\]
	Both $P$ and $\Psi$ can be constructed from $G$ in polynomial time.
	Moreover, it holds that $\sigma\models\Psi$ iff $\sigma\,x_i = \angl{v_i,c_i}$ for some
	coloring $\gamma: V\to\{1,2,3\}$ with $\gamma\,v_i=c_i$.
		It follows that $\Psi$ is satisfiable iff $G$ has a 3-coloring.
		In summary, we obtain a polynomial time reduction from the problem of 3-colorability of undirected finite graphs
	into satisfiability of finite conjunctions over some partial order.
		This concludes the proof.
	\qed.
\end{proof}

\noindent
For general partial orders $P$, however, we still may rely on the 0-normal form $\textsf{nf}_0$ and otherwise perform
the same constructions as we did for lattices with the 1-normal form.
Thus, we define an abstract ordering by
\begin{equation}
\Psi_1\sqsubseteq^\sharp \Psi_2\qquad\text{iff}\qquad
\textsf{nf}_0[\Psi_1] = \textsf{nf}_0[\Psi_1\wedge\Psi_2]
\label{def:leq}
\end{equation}
Let us denote the resulting abstract domain by $\D[P]_0$.
We have:
\begin{theorem}\label{t:general_po}
	For an arbitrary po $P$, the following holds:
	\begin{enumerate}
		\item 	If a conjunction $\Psi$ is satisfiable over $P$ then $\textsf{nf}_0[\Psi]\neq\bot$.
		\item	For all conjunctions $\Psi_1,\Psi_2$,
			$\textsf{nf}_0[\Psi_1] = \textsf{nf}_0[\Psi_1\land\Psi_2]$ implies that
			$\Psi_1\implies\Psi_2$.
	\end{enumerate}
	\qed
\end{theorem}
For arbitrary po $P$, we define the abstract projection in the same way as for conjunctions over a lattice $P$ --
only that we now rely on formulas in 0-normal form. For such a formula $\Psi$ the projection $\restrs{\Psi}{Y}$
onto a subset $Y\subseteq\X$ of variables, is again defined by removing all constraints mentioning variables not in $Y$.

It is for the abstract join operation that we must find a more general definition, since least upper bounds
or greatest lower bounds of sets of values in $P$ are no longer at hand.
Assume that $\Psi_1,\Psi_2$ are in 0-normal form and different from $\bot$. Then, we define the abstract join
$\Psi_1\sqcup^\sharp\Psi_2$ as the conjunction of the following constraints
\begin{itemize}
\item	all constraints $x\sqsubseteq y$, $x,y\in\X$, which occur both in $\Psi_1$ and $\Psi_2$;
\item	all constraints $d_i\sqsubseteq x$, $d_1,d_2\in P$, $x\in\X$ where $d_i\sqsubseteq x$ occurs in $\Psi_i$ for
	$i=1,2$ and $d_i\leq d_{3-i}$;
\item	all constraints $x\sqsubseteq d_i$, $d_1,d_2\in P$, $x\in\X$ where $x\sqsubseteq d_i$ occurs in $\Psi_i$ for
	$i=1,2$ and $d_{3-i}\leq d_i$.
\end{itemize}
This definition essentially amounts to keeping those ordering constraints between variables in which $\Psi_1$ and $\Psi_2$
agree and only keep a lower or upper bound if it is more liberal than a corresponding bound of the other formula.

\begin{example}
For the po $\Sigma^*$ with the substring ordering, consider the formulas
\[
\begin{array}{lll}
\Psi_1	&=& (\textsf{ab} \sqsubseteq x)\wedge (y\sqsubseteq \textsf{ab})\wedge(y\sqsubseteq z)		\\
\Psi_2	&=& (\textsf{abc} \sqsubseteq x)\wedge (y\sqsubseteq \textsf{abc})		\\
\end{array}
\]
Then, according to our definition,
\[
\Psi_1\sqcup^\sharp \Psi_2 = (\textsf{ab} \sqsubseteq x)\wedge (y\sqsubseteq \textsf{abc})
\]
\qed
\end{example}

\noindent
With these definitions, the binary operation $\sqcup^\sharp$ returns the least upper bound
of its arguments w.r.t.\ the ordering $\sqsubseteq^\sharp$.
Moreover, $\D[P]_0$ turns into a 2-decomposable relational domain as well.

\begin{theorem}\label{t:general_2_dec}
For every po $ P$, $\D[P]_0$ is a 2-decom\-posable relational domain.
\qed
			\end{theorem}

\subsection{\emph{Directed Domains with Disjunctions}}\label{ss:directed-dis}

Subsequently, we extend the relational domain $\D[P]$ for lattices $P$ (resp.\ $\D[P]_0$ for arbitrary po's) with disjunctions.
This extension corresponds to the \emph{disjunctive completion} of $\D[P]$ (resp.\ $\D[P]_0$) \cite{Cousot1992}.
The elements of the resulting relational domain
are disjunctions of normal form conjunctions (1-normal forms if $P$ is a lattice, and 0-normal forms in general)
where for $Y\subseteq\X$, the restriction $\restr{\Psi}{Y}$ of the disjunction $\Psi$ is defined as the disjunction of the
restrictions $\restr{c}{Y}$ of the normal form conjunctions $c$ contained in $\Psi$.
By definition, restrictions therefore are distributive.
Let $\overline\D[P]$ (resp.\ $\overline\D[P]_0$) denote the resulting relational abstract domains. If $P$ is infinite,
these relational domains have infinite strictly ascending chains, and therefore must have also strictly descending
chains of unbounded length. For the lattice $\Z$, e.g., there are even \emph{infinite} strictly descending chains, e.g.,
\[
(0\sqsubseteq x),\; (1\sqsubseteq x),\; (2\sqsubseteq x),\;\ldots
\]
Nonetheless, we have:
\begin{proposition}\label{p:2-nice}
\begin{enumerate}
\item	For every po $P$, $\overline\D[P]_0$ is 2-nice.
\item	For every lattice $P$, $\overline\D[P]$ is 2-nice.
\end{enumerate}
\end{proposition}

\begin{proof}
Let $D$ denote an arbitrary collection $\angl{d_p}_{p\in\twoClusters}$ with $d_p\in\overline\D[P]_0^p$.
Consider an arbitrary formula $d'_p$ from the set $I_{\overline\D[P]}[D]^p$.
It consists of disjunctions of conjunctions each of which may only mention variables from $p$ or constants occurring in any of the
$d_{p'}, p'\in\twoClusters$. Since the number of these formulas is finite, statement (1) follows.

The proof of the second statement is analogous -- only that the occurring constants now may also be
finite meets of constants occurring in upper-bound constraints of the initial collection or finite joins
of constants occurring in lower-boudn constraints. Still, the number of possible formulas
remains finite.
\qed
\end{proof}

\noindent
Due to \cref{p:2-nice},
the construction from
\cref{s:const} can be applied resulting in the
2-de\-com\-pos\-able relational domains $\overline\D_2^\sharp[P]$ (in case of lattices $P$) and
$\overline\D_2^\sharp[P]_0$ (for arbitrary pos).

We exemplify the construction for the lattice $\Z$ of integers, i.e., for $\overline\D_2^\sharp[\Z]$.
One-variable properties expressible in this lattice are disjunctions of interval constraints such as
\[
(x\sqsubseteq 3) \vee(5\sqsubseteq x)\wedge(x\sqsubseteq 7)
\]
Two-variable properties expressible in
this lattice are, e.g.,
\[
\begin{array}{l}
(x\sqsubseteq -1)\wedge(x\sqsubseteq y)\;\; \vee\\
(0\sqsubseteq x)\wedge(x\sqsubseteq 5)\wedge(2\sqsubseteq y)\;\; \vee\\
(6\sqsubseteq x)\wedge(y\sqsubseteq x)\wedge(y\sqsubseteq 19)
\end{array}
\]
Arbitrary elements in $\overline D_2^\sharp[\Z]$ can be understood as representations of \emph{conjunctions} of such properties.

Assume that we are given a collection $Z = \angl{s_p}_{p\in\twoClusters}$ with $s_p\in\overline\D[\Z]^p$ --
which is not yet \emph{stable}, and we would like to determine the corresponding stable collection
by performing a fixpoint iteration to determine
the greatest solution of \cref{def:constraints}.
During that iteration, we only need to consider upper and lower bounds for each variable $x$ which have already occurred in the
formulas $s_p$. Therefore, the length of each intermediate formula is bounded by a polynomial in the input, and
each unknown $r_p$ is updated only polynomially often.
As a consequence, all operations abstract join, abstract meet and abstract projection for $\overline\D_2^\sharp[\Z]$
are polynomial.
For arbitrary lattice or po $P$, we may proceed analogously. Efficiency of the fixpoint iteration, though, remains to be
checked separately for every $P$.

\subsection{\emph{Assignments}}\label{ss:directed-ass}

Let us turn to the construction of abstract transformers for assignments.
We only describe these for the relational domains $\D[P]$ and $\D[P]_0$, respectively.
We first consider three simple cases: assignments of unknown values; assignments of constants; and copying
one variable into the other.
\begin{equation}
\begin{array}{lll}
\sem{x\,{:=}\,?}^\sharp\,\Psi 	&=&	\restrs{\Psi}{\X\setminus\{x\}}	\\
\sem{x\,{:=}\,d}^\sharp\,\Psi		&=&	\restrs{\Psi}{\X\setminus\{x\}} \land (d\sqsubseteq x)\land(x\sqsubseteq d)	\\
\sem{x\,{:=}\,y}^\sharp\,\Psi		&=&	\restrs{\Psi}{\X\setminus\{x\}}
					\land(x\sqsubseteq y)\land(y\sqsubseteq x)
\end{array}
\label{def:directed_ass}
\end{equation}
for $d\in P$ and $x,y\in\X$ with $x\not\equiv y$.
Again, we realize the assignment of unknown values by restriction.
For assigning constants and variables, we remark that
equality can be expressed via a pair of inequalities.

Individual partial orders, though, may support further forms of right-hand sides in assignments.
Subsequently, we enumerate more general forms of assignments for sets and for the prefix, substring, and scattered substring
partial orders on strings.
\begin{description}
\item[\textsf{Sets.}]
	For sets, we consider right-hand sides of the form $y_1\cap y_2$
	or $y_1\cup y_2$ for $y_1,y_2 \in \X$ with $x\not\in\{y_1,y_2\}$. We define

\[
\begin{array}{lll}
	\sem{x\,{:=}\,y_1 \cap y_2}^\sharp\,\Psi		&=&
				\restrs{\Psi}{\X\setminus\{x\}} \land
				(x\sqsubseteq y_1)\land
				(x\sqsubseteq  y_2)		\\
\sem{x\,{:=}\,y_1\cup  y_2}^\sharp\,\Psi		&=&
				\restrs{\Psi}{\X\setminus\{x\}} \land
				(y_1\sqsubseteq x)\land
				(y_2\sqsubseteq x)		\\
\end{array}
\]
		Thus, we obtain after the assignment as new upper (lower) bounds of $x$ in terms of the variables $y_1$ and $y_2$.
				An analogous construction can also be applied to multisets.
				We remark that the given right-hand sides do \emph{not} entail that the equalities
		$x = y_1 \cap y_2$ and $x = y_1 \cup y_2$, respectively, hold after the assignments.
		\item[\textsf{Prefixes.}]
	In this case, right-hand sides of interest are concatenations of a constant or variable, possibly followed by
		some further value, i.e., are of the form $s\,?$ for $s$ either in $\Sigma^*$, or in $\X\setminus\{x\}$, with
		``?'' again denoting unknown input. We define
\[
\begin{array}{lll}
\sem{x\,{:=}\,s\,?}^\sharp\,\Psi		&=&	\restrs{\Psi}{\X\setminus\{x\}} \land (s\sqsubseteq x)	\\
\end{array}
\]
i.e., we only obtain information about lower bounds for $x$ after the assignment but lose all information about upper bounds.

\item[\textsf{Substrings.}]
	Again, we consider right-hand sides which are concatenations of constants or variables with
	further values. These now are of the form $?\,s_1\,?\ldots ?\,s_k\,?$
		($s_i\in\Sigma^*\cup\X\setminus\{x\}$).  We define
	\[
	\begin{array}{lll}
\sem{x\,{:=}\,?\,s_1\,?\ldots ?\,s_k\,?}^\sharp\,\Psi
					&=&	\begin{array}[t]{@{}l}
						\restrs{\Psi}{\X\setminus\{x\}}\; \land 	\\
						(s_1\sqsubseteq x)\land\ldots\land(s_k\sqsubseteq x)
						\end{array}
\end{array}
\]
For scattered substrings, we proceed similarly. In both cases, no information is obtained for upper bounds to the left-hand side variable $x$ after the assignment.
\end{description}
So far, we have assumed that the right-hand side $s$ does not contain the variable $x$ from the left-hand side.
In case that $x$ occurs in $s$,
we split the assignment into the sequence
			\[
				\textsf{tmp}\;{:=}\;s;\; x\;{:=}\;\textsf{tmp};
			\]
for some fresh variable \textsf{tmp}, i.e., first store the value of the right-hand side $s$
in \textsf{tmp} whose value only then is assigned to the left-hand side variable $x$.

These abstract tranformers for the relational domains $\D[P]$ (resp.\ $\D[P]_0$)
are readily lifted to corresponding transformers for the
weakly relational domains $\overline\D_2^\sharp[P]$ (resp.\ $\overline\D_2^\sharp[P]_0$).

\subsection{\emph{Guards and Negated Inequalities}}\label{s:noninclusions}

Let us now turn to a treatment of guards $?c$ for the directed domain $\overline\D_2^\sharp[P]$ where $P$ is a lattice.
The case for $\overline\D_2^\sharp[P]_0$ (when $P$ is not a lattice) is analogous.

A condition $c$ which consists of an inequality $s_1 \sqsubseteq s_2$ for $s_i$ being variables or constants
already represents an abstract relation. Therefore, \cref{d:guard} can be used to define the abstract effect of $\sem{?c}^\sharp$.

If the condition $c$ is a \emph{negated} inequality $s_1 \not\sqsubseteq s_2$, this is not immediately possible.
Assume that the variables occurring in $c$ all occur in $p\in\twoClusters$. Now consider an arbitrary element
$D = \angl{d_{p'}}_{p'\in\twoClusters}$. In particular, $d_p\in\overline\D[P]^p$, i.e.,
$d_p = e_1\vee\ldots\vee e_k$ for conjunctions $e_1,\ldots,e_k$ all using variables from $p$ only.
In this case, we define
\[
\begin{array}{lll}
\sem{?c}^\sharp\,D &=& D\sqcap\bigvee\{ e_j\mid e_j\not\implies(s_1\sqsubseteq s_2)\}
\end{array}
\]
Thus, the negated inequality $c$ allows to improve the abstract relation $D$ by possibly removing those conjuncts $e_j$
from $d_p$ which contradict $c$.

\section{Conclusion}\label{s:conclusion}

We considered a construction of 2-decomposable relational domains from arbitrary relational domains
and exemplified this construction by deriving 2-disjunctive constants from the relational domain of disjunctive constants.
For 2-disjunctive constants,
it turned out that normalization is prohibitively expensive.
Therefore, we provided a second general construction of 2-de\-com\-pos\-able relational domains, now based
on greatest solutions of constraint systems, which -- in the case of disjunctive constants --
results in a 2-decomposable domain where the operations join, meet, and restriction are polynomial.

In the second part, we then considered directed domains as conjunctions of inequalities over
lattices or general partial orders.
For lattices, we provided the 1-normal form for a syntactic characterization of semantic equivalence.
We showed that the resulting domain is 2-decomposable
and provided precise polynomial algorithms for 1-norma\-li\-zation, projection, join, and meet.
For arbitrary partial orders, we use a weaker form of normalization
for constructing a weaker 2-decomposable relational domain,
for which we again provided polynomial algorithms, now for 0-normalization, projection, join, and meet.
Only in the very last step, we added disjunctions by applying the general construction of 2-decomposable domain
based on approximate normalization from the previous section.
Both for 2-disjunctive constants and for directed domains, we indicated
how transfer functions for assignments and guards can be constructed.

Our results can be extended in several directions.
In the case of constants, one may, e.g., additionally, track equalities as well as disequalities between variables;
likewise for directed domains, an extensive study of the impact of negated inequalities could be of interest.
Here, we only studied lattice operations and transfer functions.
Directed domains, though, may have infinite strictly ascending chains.
Therefore, tailored widening and narrowing operators are of interest when
these domains are employed for practical static analysis.

\medskip

\paragraph{Acknowledgements.}
This work has been supported by Shota Rustaveli National Science Foundation of Georgia under the project FR-21-7973
and by Deutsche For\-schungs\-gemeinschaft (DFG) -- 378803395/2428 ConVeY.

\FloatBarrier
{
  \bibliographystyle{spmpscinat}
  \bibliography{lit}
}
\clearpage

\end{document}